\documentclass{elsarticle}
\usepackage{amsfonts}
\usepackage{amssymb}
\usepackage{amsmath,amsthm}
\usepackage{amscd}
\usepackage{color}
\makeatletter
\def\ps@pprintTitle{%
 \let\@oddhead\@empty
 \let\@evenhead\@empty
 \def\@oddfoot{}%
 \let\@evenfoot\@oddfoot}
\makeatother

\newcommand{\beq}{\begin{equation}}
\newcommand{\eeq}{\end{equation}}
\newcommand{\bea}{\begin{eqnarray}}
\newcommand{\eea}{\end{eqnarray}}
\newcommand{\beqa}{\begin{eqnarray}}
\newcommand{\eeqa}{\end{eqnarray}}
\newcommand{\nn}{\nonumber}
\newcommand\noi{\noindent}

\newtheorem{theorem}{Theorem}

\newtheorem{definition}[theorem]{Definition}

\newtheorem{proposition}[theorem]{Proposition}

\begin{document}

\begin{frontmatter}

\title{A new entropy based on a group-theoretical structure}

\author{Evaldo M. F. Curado}
\address{Centro Brasileiro de Pesquisas Fisicas and National Institute of Science and Technology for Complex Systems, Rua Xavier Sigaud 150, 22290--180 Rio de Janeiro -- RJ, Brazil}
\ead{evaldo@cbpf.br}
\author{Piergiulio Tempesta}
\address{Departamento de F\'{i}sica Te\'{o}rica II, M\'{e}todos y modelos matem\'{a}ticos, Facultad de F\'{\i}sicas, Universidad Complutense, 28040 -- Madrid, Spain and Instituto de Ciencias Matem\'aticas, C/ Nicol\'as Cabrera, No 13--15, 28049 Madrid, Spain}
\ead{p.tempesta@fis.ucm.es, piergiulio.tempesta@icmat.es}
\author{Constantino Tsallis}
\address{Centro Brasileiro de Pesquisas Fisicas and National Institute of Science and Technology for Complex Systems, Rua Xavier Sigaud 150, 22290--180 Rio de Janeiro -- RJ, Brazil, and Santa Fe Institute, 1399 Hyde Park Road, Santa Fe, NM 87501, USA}
\ead{tsallis@cbpf.br}

\begin{abstract}
A multi-parametric version of the nonadditive entropy $S_{q}$ is introduced. This new entropic form, denoted by  $S_{a,b,r}$, possesses many interesting statistical properties, and it reduces to the entropy $S_q$ for $b=0$, $a=r:=1-q$ (hence Boltzmann-Gibbs entropy $S_{BG}$ for $b=0$, $a=r \to 0$). The construction of the entropy $S_{a,b,r}$ is based on a general group-theoretical approach recently proposed by one of us \cite{Tempesta2}. Indeed, essentially all the properties of this new entropy are obtained as a consequence of the existence of a rational group law, which expresses the 
structure of $S_{a,b,r}$ with respect to the composition of statistically independent subsystems. Depending on the choice of the parameters, the entropy $S_{a,b,r}$ can be used to cover a wide range of physical situations, in which the 
measure of the accessible phase space increases say exponentially with
the number of particles $N$ of the system, or even stabilizes, by increasing $N$, to a limiting value. This paves the way to the use of this entropy in contexts where a system "freezes" some or many of its degrees of freedom by increasing the number of its constituting particles or subsystems.
\end{abstract}

\date{April 27, 2015}

\end{frontmatter}
\newpage
\tableofcontents

\section{Introduction}

In the last decades, the non-extensive scenario, originally proposed in \cite{Tsallis1}, has been largely investigated as a new thermodynamic framework allowing to generalize the standard Boltzmann-Gibbs approach to new physical contexts, where ergodicity hypothesis is violated \cite{Tbook}. At the same time, the ubiquity of the notion of entropy in social sciences paved the way to fruitful extensions of the standard information theory of  Shannon and Khinchin  \cite{Shannon, Shannon2, Khinchin} towards new classical and quantum formulations.

The search for new entropic forms has been very active in the last decades. Many different entropies have been proposed, generalizing the Boltzmann-Gibbs entropy from different perspectives (see, e.g., \cite{Abe, Beck, Hanel, GellMann, Kaniad1, Naudts, Tempesta1, TsallisCirto}.

In particular, a group-theoretical approach to the notion of entropy has been advocated in  \cite{Tempesta2}. It is based on the observation that a thermodynamically admissible entropy should satisfy not only the
first three Khinchin axioms (continuity, concavity, expansibility), but also a general \textit{composability property}. It amounts to require that, given an entropic functional $S$, its values on a system defined by the union of two statistically independent subsystems $A$ and $B$ should depend (in addition to a possible small set of fixed indices; for instance the index $q$ for $S_q$) on the entropies of the two subsystems only . This property can be imposed on full generality (and we shall talk about \textit{strict composability}) or at least on subsystems characterized by the uniform distribution (composability in a \textit{weak sense} or composability tout court). This last property applies, for instance, when considering isolated physical systems at the equilibrium (microcanonical ensemble), 
or in contact with a thermostat 
at very high temperature 
(canonical ensemble) \footnote{To be more precise, let us illustrate the BG case.  The formula $S_{BG}= \ln W$ (assuming $W$ to be finite) applies to both microcanonical and $T\to\infty$ canonical cases, but, in the former, $W$ refers to the total number of states within a thin slice of phase space corresponding to a given total energy, whereas, in the latter, $W$ refers to  the total number of states within the entire phase space.
}. 
In full generality, it amounts to say that there exists a continuous function of two real variables $\Phi(x,y)$ such that
\text{(C1)}
 \beq
S(A \cup B)=\Phi(S(A),S(B);\{\eta\}) \label{C1}
\eeq
where $\{\eta\}$ is a possible set of real continuous parameters, and
$A\subset X$ and $B\subset X$ are two statistically independent subsystems of a given system $X$, with the further properties

(C2) Symmetry:
\beq
\Phi(x,y)=\Phi(y,x). \label{C2}
\eeq

(C3) Associativity:
\beq
\Phi(x,\Phi(y,z))=\Phi(\Phi(x,y),z). \label{C4}
\eeq 

(C4) Null-composability:
\beq
\Phi(x,0)=x. \label{C3}
\eeq
The
Boltzmann-Gibbs, the $S_q$, the R\'enyi  entropies are known to be strictly composable. Instead, the weak composability property is shared by infinitely many more entropies. A huge class of admissible entropies (more precisely, continuous, concave, expansible, composable) is provided by the \textit{universal-group entropy}, related with the Lazard universal formal group. This class possesses many remarkable properties, in particular a Legendre structure, extensivity in suitable regimes, Lesche stability in suitable regimes, among others. The two most frequent examples of entropies of this class, i.e. possessing a group-theoretical structure, are Boltzmann-Gibbs entropy and $S_q$ entropy. They have associated the additive group law 
\beq
S(A \cup B)= S(A)+ S(B) \label{additive}
\eeq
and the multiplicative one \cite{Haze}
\beq
S(A \cup B)= S(A)+ S(B)+a S(A) S(B), \label{multiplicative}
\eeq
respectively. In the literature, $a\in\mathbb{R}$ is usually written in the form $a := 1-q$.

In this paper, we explore a remarkable example of \textit{rational group law}: 
\beq
S(A \cup B)= \frac{S(A)+ S(B)+a S(A) S(B)}{1+b S(A) S(B)} \label{rational} \, ,
\eeq
where $a,b\in\mathbb{R}$. The corresponding rational group law is given by 
\beq
\Phi(x,y) = \frac{x+y+a xy}{1+b xy} \label{philaw}
\eeq
Notice that when $a=b=0$, we recover the standard additive law \eqref{additive}; for $b=0$, we recover the case \eqref{multiplicative}.

It is interesting to notice that for $b=0$ this relation can be written in an additive form, namely 
\begin{equation} 
\label{additive1}
\frac{\ln[1+a \Phi(x,y)]}{a} = \frac{\ln[1+ a x]}{a} + \frac{\ln[1+ a y]}{a} \, ,
\end{equation}
and this yields to R\'enyi entropy. 
Whenever $b\neq0$, we have a genuinely new case.  Notice  that, for the particular instances    
$b=a+1 \neq 0$ and $b = 1-a \neq 0$, 
the following  additive rules hold respectively: 
\begin{align} 
\label{additive2}
\ln\left[\frac{1-\Phi(x,y)}{1+(a+1)\Phi(x,y)}\right] &= \ln\left[\frac{1-x}{1+(a+1)x}\right] + \ln\left[\frac{1-y}{1+(a+1)y}\right] \\
\label{additive3}
\ln\left[\frac{1+\Phi(x,y)}{1+(a-1)\Phi(x,y)}\right] &= \ln\left[\frac{1+x}{1+(a-1)x}\right] + \ln\left[\frac{1+y}{1+(a-1)y}\right] \, .
\end{align}
These two relations can be interchanged one into the other through the transformation $(x,a) \rightarrow (-x,-a) $.
It might be interesting to explore these properties in future works.

At the best of our knowledge, this two-parametric rational group law was not previously considered in the literature. A specific one-parametric realization of it, i.e
\beq
\Phi(x,y) = \frac{x+y+(\alpha-1) xy}{1+\alpha xy} \label{philaw2}
\eeq
plays an important role in algebraic topology. Precisely, for $\alpha=-1,0,1$, we obtain group laws respectively associated with the Euler characteristic, the Todd genus and the Hirzebruch $L$-genus \cite{BMN}.
Notice that identifying $a \equiv \alpha-1 $ and $b \equiv \alpha$,    
Eq. (\ref{philaw}) turns out to be Eq. (\ref{philaw2}). Also, these relations between $\alpha$, $a$ and $b$ are exactly the necessary conditions to get the additive property given in Eq. (\ref{additive2}).

The aim of our work is to construct the entropy associated with the rational group law \eqref{rational}. It is easy to verify that it satisfies eqs. \eqref{C2}-\eqref{C3}. Also, it admits an inverse, i.e. there exists a real function $\phi(x)$ such that
\beq
\Phi(x,\phi(x))=0.
\eeq
For instance, for the case of the additive law, $\phi(x)=-x$. 
For the rational group law \eqref{rational}
we have  $\phi(x)=-x/(1+ax)$. This is the reason why we talk about the group structure associated with an entropy \cite{Tempesta2}. Notice however that, if we restrict to values $x,y\in\mathbb{R}^{+}\cup \{0\}$ (as mandatory for physically reasonable entropic forms), it is not possible to define an inverse.

In the following, we shall construct explicitly the entropy associated with the law \eqref{rational}.

\section{The new entropy and its properties}

\begin{definition} The $S_{a,b,r}^{(+)}$ entropy, for $r>0$, is the function
\beq
S_{a,b,r}^{(+)}[p] =\sum_{i=1}^{W} s^{(+)} [p_i]:=
\sum_{i=1}^{W} p_i Log_{a,b,r}^{(+)}\left(\frac{1}{p_i}\right) \label{newentplus} \, ,
\eeq
where the generalized {\it plus logarithm}  is defined as
 \beq
Log_{a,b,r}^{(+)}(x):= \frac{2 (x^{r}-1)}{-a(x^{r}-1) + \sqrt{a^2+4b}\hspace{1mm}(x^{r}+1) }\, . \label{genlogplus}
\eeq 
The $S_{a,b,r}^{(-)}$ entropy, for $r<0$, is the function
\beq
S_{a,b,r}^{(-)}[p]:= \sum_{i=1}^{W} s^{(-)} [p_i]:=
 \sum_{i=1}^{W} p_i Log_{a,b,r}^{(-)}\left(\frac{1}{p_i}\right) \label{newentminus} \, ,
\eeq
 where the generalized {\it minus logarithm} is given by
\beq
Log_{a,b,r}^{(-)}(x):= \frac{2 (x^{r}-1)}{-a(x^{r}-1) - \sqrt{a^2+4b}\hspace{1mm}(x^{r}+1) }. \label{genlogminus}
\eeq 
Here $a,b$ are real parameters.
\end{definition}

\noi These entropies satisfy many remarkable properties.
\begin{proposition} The entropies $S_{a,b,r}^{(\pm)}$ reproduce the standard $S_{BG}$ entropy for $b=0$,  $a=r$, in the limit $r \to 0$:
\beq 
\lim_{r\to 0} S_{r,0,r}^{(\pm)}[p]=S_{BG}[p]. 
\eeq
\end{proposition}
Proof. Indeed, we have that
 \beq
S_{a,0,r}^{(\pm)}[p]= \sum_{i=1}^{W}p_i \frac{  p_i^{-|r|}-1}   {a} \, ,
\eeq
valid for any value of $r$.
Therefore, for $a=r$ and taking the limit $r\to 0$, the previous expression converts into the entropy $S_{BG}$.
\begin{proposition}
The  entropies
\eqref{newentplus}
and \eqref{newentminus} satisfy the first three  Khinchin axioms.
\end{proposition}
Indeed,  they are \textit{continuous function} of all their arguments. Second,
the function $s^{(+)} [p]$,
defined in Eq. \eqref{newentplus} for $r>0$,
is concave, ensuring that the maximum of the entropy is attained at equiprobability,
 if the following conditions are satisfied for $W\geq 1$:
\begin{align}
\label{conditionplus}
i) & \,\,\, 0<r \leq 1 & \rightarrow  & \,\,a <  0  \,\,\,\mbox{and} \,\,\, b>  \frac{-a^2}{4} \\
                    &     &         & a \geq 0 \,\,\, \mbox{and} \,\,\, b > 0 \\
ii) & \,\,\,\,\,\, \,\,\, r > 1 &  \rightarrow &  \,\,a \leq 0  \,\,\,\mbox{and} \,\,\, b > \frac{-a^2}{4} \\
                      &           &                &  \,\, a > 0  \,\,\,\mbox{and} \,\,\, b > \frac{a^2 (r^2-1)}{4}
\end{align}

Also, the function $s^{(-)} [p]$, defined in Eq. \eqref{newentminus} for $r<0$, is concave if the following conditions are verified for $W\geq 1$:
\begin{align}
\label{conditionminus}
i) & \,\,\, -1 \leq r < 0 & \rightarrow &  \,\,a < 0  \,\,\,\mbox{and} \,\,\,  b > \frac{-a^2}{4}   \\
   &                    &                   &   \,\, a \geq 0   \,\,\,\mbox{and} \,\,\,  b > 0 \\
ii) & \,\,\,\,\,\, \,\,\, r< -1   &  \rightarrow  & \,\,a \leq 0  \,\,\,\mbox{and} \,\,\, b > \frac{-a^2 }{4}\\
                      &          &               &   \,\, a > 0  \,\,\,\mbox{and} \,\,\,  b > \frac{a^2 (r^2-1)}{4}
\end{align}

Third, it is also possible to show that
$p Log_{a,b,r}^{(\pm)}\left(1/p \right)  \rightarrow 0$ when $p \rightarrow 0$, 
which ensures expansibility.  Also, 
$Log_{a,b,r}^{(\pm) } (1) = 0$, yielding the value zero for the entropy when we 
have certainty. 

\begin{proposition}
The generalized logarithms \eqref{genlogplus} and \eqref{genlogminus}
satisfy the group law \eqref{rational} and
have the properties: 
 \begin{align}
Log_{a,b,r}^{(+)}[p] &= Log_{a,b,-r}^{(-)}[p] \label{proplog1}, \\
Log_{a,b,-r}^{(\pm)}[p] &= Log_{a,b,r}^{(\pm)}[1/p] , \\
Log_{a,b,r}^{(+)}[p] & = - Log_{-a,b,r}^{(-)}[p]
\end{align}
implying that $Log_{a,b,r}^{(\pm)}[p]  = -Log_{-a,b,-r}^{(\pm)}[p]$. 
\end{proposition}
These equalities can be proven by direct calculations. 
These properties enable us to say that the  entropies  \eqref{newentplus} and \eqref{newentminus} are \textit{thermodynamically admissible} \cite{Tempesta2}. It means that our entropy satisfies the first three SK axioms and in addition the generalized logarithm is composable in the sense of eqs. (1)-(4).
  
\begin{figure}
\begin{center}
\includegraphics[width=4.2in]{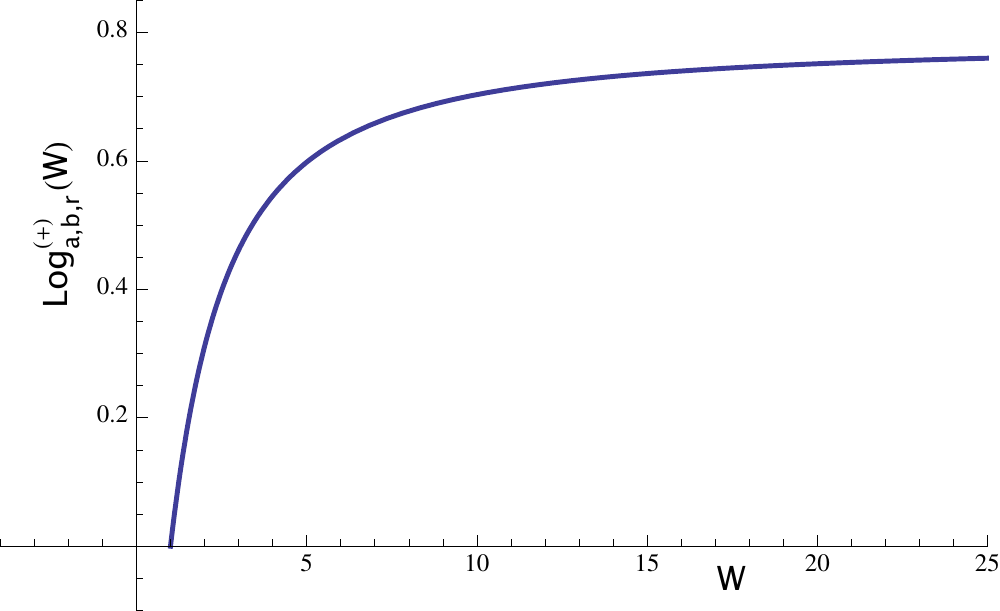}
\caption{$Log_{a,b,r}^{(+)}$ in function of $W$ for $a=0.3$, $b=2$ and $r=1.3$.}
\label{figlog}
\end{center}
\end{figure}

Constrained to the regions allowed in 
Proposition 3, the generalized logarithms  $Log_{a,b,r}^{(\pm)}(W)$ are monotonically increasing functions of $W$, see Fig. \ref{figlog}. The graphics for the other allowed regions, shown in Proposition 3, have qualitatively the same behavior.  
 
\section{The construction of the entropy $S_{a,b,r}$ from the rational group law}

In this section, we solve in full generality the following inverse problem: \textit{given the group law \eqref{rational}, find the associated entropy}.
Therefore, according to the general framework proposed in \cite{Tempesta2}, assume that there exists a continuous function
\beq
\Phi(x,y)=x+y+ \text{higher order terms} \label{phigen}
\eeq
that satisfies the group properties \eqref{C2}-\eqref{C3}. Notice that the rational law \eqref{philaw} is of this form.
\noi  The  entropy associated with \eqref{phigen} can be constructed as follows. We look for a function $G(t)$, which a priori is a formal power series, such that
\beq
\Phi(x,y)=G(G^{-1}(x)+G^{-1}(y)) \label{Lazard}.
\eeq
This function $G(t)$ is called the formal group exponential \cite{Haze}. Here $G^{-1}(s)$ is the compositional inverse of $G(t)$, i.e. $G(G^{-1}(s))=s$ and $G^{-1}(G(t))=t$. The most general form of $G(t)$ is
\beq
G(t)=\sum_{k=0}^{\infty} A_{k} \frac{t^{k+1}}{k+1}= A_{0} t+ A_{1} \frac{t^2}{2}+A_{2} \frac{t^3}{3}+\ldots. \label{G}
\eeq
When considering this general form, equation \eqref{Lazard} defines the Lazard universal formal group \cite{Haze}.
The entropy associated with \eqref{phigen} is unique if we fix $A_{0}=1$, and will be of the form
\beq
S_{U}[p]:=\sum_{i=1}^{W} p_i G\left(\ln \frac{1}{p_i}\right) \label{UGE},
\eeq
i.e. will be a representative of the \textit{universal-group entropy}.
The inverse $G^{-1}(s)$ can be computed by means of the Lagrange inversion theorem. We get:
\beq
G^{-1}(s)=\frac{s}{A_{0}}-\frac{A_1}{2 (A_0)^3}s^2+
\eeq
The case of the BG entropy is immediately obtained. In this case $\Phi(x,y)=x+y$, and the Lazard law gives $G(t)=t$. From eq. \eqref{UGE} we get back the BG case.

Let us apply the previous theory to the case of interest, i.e. the rational group law \eqref{philaw}. Now we have that $A_{k}=A_{k}(a,b)$. The expansion of the function $\Phi(x,y)$ is
\bea
\Phi(x,y)&=& x+y + a x y -b (xy^2+y x^2)- a b x^2 y^2 \nn \\ \nn &+&b^2 (x^2 y^3+x^3 y^2)+ a b^2 x^3 y^3 \\
&+& \text{higher order terms} \label{HOT}
\eea
By using the expression \eqref{Lazard} with the form \eqref{G} for the expansion of the formal exponential, and identifying the terms appearing in this expansion with those coming from \eqref{HOT}, we get
an infinite set of relations for the coefficients $A_k$:
\bea \label{coefficients}
A_{0} &\in&\mathbb{R} \nn \\
A_{1} &=& a A_{0}^2 \nn \\
A_2&=& \frac{1}{2}\left(a^2 - 2 b\right)A_{0}^3 \nn \\
A_3&=& \frac{1}{3!}\left(a^3 - 8 a b\right)A_{0}^4 \nn \\
A_4&=& \frac{1}{4!}\left(a^4 - 22 \hspace{1mm} a^2 b + 16 \hspace{1mm} b^2\right)A_{0}^5 \nn \\
A_5&=& \frac{1}{5!}\left(a^5 - 52 \hspace{1mm} a^3 b+136\hspace{1mm} a b^2 \right)A_{0}^6 \nn \\
A_6&=& \frac{1}{6!}\left(a^6 - 114 \hspace{1mm} a^4 b+ 720 \hspace{1mm} a^2 b^2 - 272\hspace{1mm} b^3 \right)A_{0}^7 \\
...
\eea
A priori, the coefficients $A_{k}$ provide the most general solution to our problem.
Before proceeding further, let us consider the particular case $b=0$. Then
\beq
\Phi(x,y)= x+y+ a xy
\eeq
If we put $b=0$ in the previous coefficients \eqref{coefficients}, and let $A_{0}=1$, we get immediately
\beq
a_{k}= \frac{1}{k!} a^{k}
\eeq
i.e.
\beq
G(t)=\frac{e^{at}-1}{a},
\eeq
which is the correct form we were looking for. Indeed, according to the prescription \eqref{UGE}, and putting $a=1-q$  we get back the $S_q$ entropy
\beq
S=\sum_{i=1}^{W} p_i \frac{p_i^{q-1}-1}{1-q}=\frac{1-\sum_{i=1}^{W}  p_{i}^{q}}{q-1} .
\eeq
The general case provides us with a series solution: indeed, we reconstruct $G(t)$ and hence the entropy \eqref{UGE}, term by term.  However, this series can be re-summed, to give
the \textit{closed form solution}

\beq
G(t)=\frac{2 (e^{r t}-1)}{-a(e^{r t}-1)\pm\sqrt{a^2+4b}(e^{r t}+1) } \label{finalG}
\eeq 
In particular it emerges that the arbitrary coefficient is fixed to be
\beq
A_{0}=\pm \frac{r}{\sqrt{a^2+4b}}.
\eeq
The realization of the universal-group entropy $\sum_{i}p_i G\left(\ln \frac{1}{p_i}\right)$ for $G(t)$ given by eq. \eqref{finalG} is nothing but the entropies \eqref{newentplus} and \eqref{newentminus}. 

The solution  in closed form \eqref{finalG}, or equivalently the generalized logarithms \eqref{genlogplus} and \eqref{genlogminus}, can also be obtained from the group law \eqref{philaw} by a direct procedure. It entails the formulation
of a suitable ansatz for the solutions of \eqref{philaw}.

The previous approach has the advantage of being systematic, i.e. it can be used in full generality from any group law of the form \eqref{phigen}, and does not demand any guess or ansatz.
This method amounts to the construction of a series solution for the functional equation
\eqref{rational}, and is conceptually similar to the technique for generating series solutions of differential equations.

\section{On the extensivity of the entropy $S_{a,b,r}$}

One of the main reasons to consider generalized entropies is the fact that they can  
 be useful, or even mandatory, to describe systems with unusual behavior. 
If an entropy is extensive, it essentially 
means that, for an occupation law $W=W(N)$ of the phase space associated with a given system, it is asymptotically proportional to $N$, the number of particles of the system. Precisely, $S_{BG}$ is extensive whenever $W(N)\sim k^{N}$, where $k\in\mathbb{R}_{+}$ is a suitable constant. However, for substantially different choices of $W=W(N)$, this property is no longer true for $S_{BG}$.

A natural question is to ascertain whether the new entropy we propose in this paper \textit{is} extensive. Its group structure, once again, ensures that this property holds for a suitable asymptotic occupation law $W=W(N)$ of phase space.
A general result proven in \cite{Tempesta2} is that a sufficient condition for an entropy of the form \eqref{UGE} to be extensive is that
\beq
\ln W(N)\sim G^{-1}(N), \label{asymptoccuplaw}
\eeq 
provided that the real function $W(N)$ be defined for all $N\in\mathbb{N}$, with 

\noi $\lim_{N\to\infty} W(N)=\infty$. These requirements usually restrict the space of parameters.
In our case, 
we observe that when $p_i=1/W$ for all $i=1,\cdots, W$, the entropies \eqref{newentplus}, for $r>0$, and \eqref{newentminus}, for $r<0$
 (or, more generally, the entropy given by eq. (\ref{UGE})),
tend to the limit value
\bea
\label{winf}
S_{a,b,r}^{(\pm)}[1/W] \rightarrow \frac{2}{\sqrt{a^2+4 b} -a }  \, , 
\eea          
if $\lim_{N\to\infty} W(N)=\infty$. In particular, for $b\to 0$, the entropy diverges; as a consequence of the previous discussion, there exists a regime of extensivity, for $W(N)\sim N^{\gamma}$, with $\gamma=\frac{1}{a}$.

\noi If $\lim_{N\to\infty} W(N)=c\in \mathbb{R_+}$, with $c>1$,  these entropies tend to the value 

\bea
\label{winf2}  
S_{a,b,r}^{(\pm)}[1/c] \rightarrow \frac{2 \left(c^r-1\right)}{\pm \sqrt{a^2+4 b} \left(c^r+1\right)-a \left(c^r-1\right)}            \, .
\eea

In both cases, if $b \neq 0$ the limiting value is finite, independently on whether $W$ diverges or tends to a constant for $N\to\infty$. Consequently, the $b \ne 0$ entropies, albeit monotonically increasing functions of $W$, can not be extensive. 

A natural question emerges, concerning the kind of systems such type of entropies could be useful for. A possible answer is that 
the present formalism
could be relevant whenever treating systems highly connected, where the addition of a new degree of freedom essentially does not change the value of the entropy, for a large number of degrees of freedom. For example, if we add a molecule of water in a glass of liquid water, after enough time the molecule will move everywhere inside the glass. However, if we release a molecule of water in a glass filled with ice,  the additional molecule will eventually freeze. The increase of the entropy value is substantially lower in the latter case than in the former. 

One can also consider different scenarios, borrowed from social sciences, where no thermodynamical or energetical aspects are involved, and extensivity is a priori  not required. Again such entropies, that increase very little with the addition of new degrees of freedom, could be play a relevant role in describing situations where the amount of information tends to stabilize, irrespectively of the increase of new agents involved in the information exchange.

Consequently, the multiparametric entropy $S_{a,b,r}$ is compatible with both scenarios: the standard one, where an increase of the numbers of degrees of freedom converts into an increase of the entropy, and the ``anomalous" one, where an increase of the number of particles essentially freezes the system, by confining it in the phase space.
Excepting for $S_q$ for $q>1$ and $S_\delta$ for $\delta<0$ (see \cite{TsallisCirto} and references therein), this flexibility in the limit $W\to\infty$ is seemingly not shared by the entropies typically used in the literature.

\section*{Acknowledgments}
The research of P. T. has been partly supported by the research project FIS2011--22566, Ministerio de Ciencia e Innovaci\'{o}n, Spain. Partial financial support by CNPq and Faperj (Brazilian agencies), and by the John Templeton Foundation is acknowledged as well.

\vspace{3mm}

\noi 

\end{document}